\documentclass[a4paper,twocolumn,showpacs,superscriptaddress,floatfix]{revtex4-1}

\usepackage{latexsym}
\usepackage{amssymb}
\usepackage{amsfonts}
\usepackage{amsmath}
\usepackage{bm}
\usepackage{graphicx}   
\usepackage{color}
\usepackage{subfigure}
\usepackage{times}
\usepackage{units}
\usepackage{hyperref}

\begin{document}


\title{Magnetorotational instability in relativistic hypermassive
  neutron stars}

\author{Daniel M. Siegel}
\author{Riccardo Ciolfi}
\author{Abraham I. Harte}
\author{Luciano Rezzolla}

\affiliation{%
 Max Planck Institute for Gravitational Physics (Albert Einstein Institute),\\
 Am M\"uhlenberg 1, 14476 Potsdam-Golm, Germany
}

\date{\today}

\begin{abstract}
A differentially rotating hypermassive neutron star (HMNS) is a
meta\-stable object which can be formed in the merger of neutron-star
binaries. The eventual collapse of the HMNS into a black hole is a key
element in generating the physical conditions expected to accompany the
launch of a short gamma-ray burst. We investigate the influence of
magnetic fields on HMNSs by performing three-dimensional simulations in
general-relativistic magnetohydrodynamics. In particular, we provide
direct evidence for the occurrence of the magnetorotational instability
(MRI) in HMNS interiors. For the first time in simulations of these
systems, rapidly-growing and spatially-periodic structures are observed
to form with features like those of the channel flows produced by the MRI
in other systems. Moreover, the growth time and wavelength of the
fastest-growing mode are extracted and compared successfully with
analytical predictions. The MRI emerges as an important mechanism to
amplify magnetic fields over the lifetime of the HMNS, whose collapse to
a black hole is accelerated. The evidence provided here that the MRI can
actually develop in HMNSs could have a profound impact on the outcome of
the merger of neutron-star binaries and on its connection to short
gamma-ray bursts.
\end{abstract}


\maketitle


\section{Introduction}
\vspace{-0.3cm}
The magnetorotational instability (MRI)
refers to exponentially growing modes that can develop in differentially
rotating magnetized fluids \cite{Velikhov1959,*Chandrasekhar1960,
  *Balbus1991}, and is believed to play a pivotal role in a variety of
astrophysical systems. Various analytic and numerical studies agree that
through the generation of turbulence, the MRI is the main mechanism for
the outward transport of angular momentum in accretion disks around
compact objects \cite{BalbusHawley1998,*Balbus2003,*Armitage2011}. The
MRI can also play a role in core-collapse supernovae, either by powering
the explosion through the conversion of rotational energy into magnetic
energy and the production of a magnetohydrodynamic (MHD) outflow
\cite{Akiyama2003, *Burrows07a}, or as a source of thermal energy
generated by the MRI-induced turbulence and adding to a neutrino-driven
explosion \cite{Thompson2005, Janka12}. MRI effects are
particularly important when modelling high-energy supernovae and
hypernovae \cite{Janka12}.

Here we consider a further scenario where the MRI may play a crucial
role: the evolution of hypermassive neutron stars (HMNSs). HMNSs are
metastable objects that can be formed by the merger of neutron star
binaries \cite{Shibata05c,Baiotti08}. They are differentially-rotating
neutron stars which exceed the mass limits of rigidly rotating stars. The
eventual collapse of a HMNS---induced either by neutrino
cooling~\cite{Paschalidis2012}, or by the removal of differential
rotation via magnetic fields \cite{Giacomazzo:2010}, fluid viscosity, or
gravitational radiation \cite{Baiotti08}---leads to a spinning black
hole surrounded by a hot and dense torus (see~\cite{Shibata06a,
  *Rezzolla:2010} for a discussion). The evolution of magnetic fields in
HMNSs is of great importance since their rearrangement following
amplification by magnetic winding and the MRI may provide the necessary
conditions to launch the relativistic jets observed in short gamma-ray
bursts (SGRBs) \cite{Duez:2006qe, Giacomazzo:2010, Rezzolla:2011,
  Shibata06b,*Kiuchi2012b}.

Numerical simulations of the MRI face a fundamental challenge: The
wavelength of the fastest growing mode of the instability is proportional
to the magnetic field strength and is typically much smaller than the
scale of the astrophysical system considered. Due to computational
limitations, many simulations, therefore, fail to resolve the MRI unless
very high initial magnetic fields are employed, or only a small part of
the system is simulated as in local or semi-global simulations
(e.g., \cite{Hawley92,Obergaulinger:2009,Masada2012}), or the number of
spatial dimensions is reduced via symmetries (e.g., \cite{Duez2006a,
  Duez:2006qe, Shibata06c, *Obergaulinger06, *Obergaulinger06b,
  Cerda2008}). In addition, most simulations attempting to resolve the
MRI were conducted within Newtonian or special-relativistic MHD. The most
advanced general-relativistic results on the MRI in HMNSs date back to
the exhaustive work of \cite{Duez:2006qe,Duez2006a}, where the system was
studied in axisymmetry and a specific stage of the magnetic-field
amplification was interpreted as evidence for the MRI.

Here, we focus on the pre-collapse phase of the HMNS evolution and
provide evidence for the occurrence of the MRI in global,
three-dimensional and fully general-relativistic MHD simulations. The
emergence of well-resolved coherent channel flows allows us to measure
quantities such as the wavelength and the growth rate of the fastest
growing mode, opening the way to a systematic study of the MRI in HMNSs.

\vspace{-0.2cm}
\section{Numerical setup}
\vspace{-0.3cm}
As a typical HMNS, we consider the
axisymmetric initial model A2 of \cite{Giacomazzo2011}, which is
constructed using the \texttt{RNS} code \cite{Stergioulas95}. This
assumes a polytropic equation of state (EOS) $p=K\rho^{\Gamma}$, where
$p$ denotes the fluid pressure and $\rho$ the rest-mass density, with
$K=100$ (in units where $c=G=M_\odot=1$) and $\Gamma=2$. The initial HMNS
has an Arnowitt-Deser-Misner (ADM) mass of $M=2.23\, M_\odot$ and is differentially rotating
according to a ``$j$-constant law'' with central angular velocity
$\Omega_c=(u^\phi/u^t)_c=2\pi\times 7.0\;\textrm{kHz}$, where $u^\mu$ is
the fluid 4-velocity. On top of this purely hydrodynamic equilibrium
model, we add a poloidal magnetic field confined inside the star and
specified by the vector potential $A_\phi = A_b \varpi^2 \mathrm{max}
\{(p-0.04\,p_\textrm{max}),0\}$, where $\varpi$ denotes the cylindrical
radius and $p_\textrm{max}$ the maximum fluid
pressure~\cite{Giacomazzo:2010}. We tune $A_b$ so as to have central
magnetic fields $B_c^\mathrm{in}=(1-5)\times10^{17}$~G. Despite the very
high resolutions employed here, such strong magnetic fields are essential
to resolve the MRI. Even at these strengths, however, the average
magnetic-to-fluid pressure ratios in these models are only
$(0.045-1.2)\times10^{-2}$.

Our simulations are performed with the \texttt{Whisky}
\cite{Giacomazzo:2007ti} and the \texttt{Ccatie}
codes~\cite{Pollney:2007ss}. These solve the coupled Einstein-MHD
equations in 3+1 dimensions on a Cartesian grid employing high-resolution
shock-capturing schemes and the conformal traceless decomposition of the
ADM formulation of the Einstein equations (see~\cite{Giacomazzo:2010} for
details). The fluid is assumed to follow ideal MHD and the ideal-fluid
EOS $p=(\Gamma -1)\rho\epsilon$, where $\epsilon$ is the specific
internal energy and $\Gamma=2$. The computational grid comprises a
spatial box of dimensions $[0, 94.6]\times [0, 94.6]\times [0, 53.9]$ km
with four mesh-refinement levels~\cite{Schnetter-etal-03b} and a fiducial
finest resolution with coordinate grid spacing $h = 44\,\textrm{m}$. This
is comparable to the $h \simeq 37\,\textrm{m}$ used in previous
simulations (which, however, assumed axisymmetry)
\cite{Duez:2006qe,Duez2006a}. All of the results presented here refer to
the finest refinement level, which corresponds to a spatial domain of $[0, 11.8]\times [0, 11.8]\times [0, 6.7]$ km
and thus fully covers the HMNS at all times. To
reach high enough spatial resolutions and make these calculations
possible at all, we employ a reflection symmetry across the $z=0$ plane
and a $\pi/2$ rotation symmetry around the $z$-axis. Repeating some
simulations with $\pi$ symmetry does not alter results found by assuming
$\pi/2$ symmetry. The $z$-symmetry provides large computational savings,
but suppresses the toroidal field in the equatorial plane. However,
the instability develops far from the equatorial plane and thus the
onset and early evolution of the instability are unlikely to be affected by such symmetry.

\vspace{-0.2cm}
\section{Analytical predictions}
\vspace{-0.3cm}
In our initial axisymmetric
configuration, magnetic fields are purely poloidal while the fluid
velocity is purely toroidal. The fluid does not rotate uniformly along
magnetic field lines, so the magnetic field is ``wound up" as the HMNS
rotates. Assuming axisymmetry and a sufficiently slow variation of the
3-metric $\gamma_{ij}$, of the poloidal magnetic field and of the angular
velocity, the induction equation can be used to show that
(cf., e.g., \cite{Duez:2006qe})
\begin{equation}
B_{\mathrm{tor}} \approx ( \varpi B^i \partial_i \Omega ) t =
a_\mathrm{w} t\,.
\label{eq:winding}
\end{equation}
Here, $B_{\mathrm{tor}} = \gamma_{ij} B^i e^j_\phi$, where $e_\phi^i$ is
the unit vector proportional to the azimuthal Killing field. The
linear-in-time growth is expected only during the first phase of the
evolution.

There exists no adequate theoretical description of the MRI in systems of
the type considered here. Nevertheless, we observe effects similar to
those known to arise in simpler systems like accretion disks. In
particular, certain short-wavelength modes appear to be preferentially
amplified over time. From a linear perturbation analysis of the Newtonian
MHD equations for axisymmetric perturbations, which can at best hold
approximately in our system, the characteristic timescale and wavelength
for the fastest growing mode with wave vector $k_{_\mathrm{MRI}}^i$ may
be estimated by (see \cite{BalbusHawley1998} and, e.g.,
\cite{Duez:2006qe})
\begin{eqnarray}
\tau_{_\mathrm{MRI}} \sim \Omega^{-1}\,, \qquad
\lambda_{_\mathrm{MRI}} \sim
\left(\frac{2\pi}{\Omega}\right) 
\left(\frac{B_i e^i_k}{\sqrt{4\pi\rho}}\right)\,
\label{eq:tau_lambda}
\end{eqnarray}
on an order-of-magnitude level, where $e_k^i$ is the unit vector along
$k_{_\mathrm{MRI}}^i$. Note that $\tau_{_\mathrm{MRI}}$ is independent of
$B$ while $\lambda_{_\mathrm{MRI}}$ is linear in it. If these estimates
are approximately valid for our system, they can only be expected to hold
in an appropriate ``inertial frame.'' As the 4-metric $g_{\mu\nu}$ in
the singularity-avoiding coordinates of our simulations is
significantly different from the flat spacetime one, the estimate for
$\lambda_{_\mathrm{MRI}}$ needs to be corrected by multiplying it by a
factor $\sqrt{-g_{00}}$ (which can be quite far from unity). Ignoring
this correction can easily lead to inappropriate estimates for the
numerical resolution required to resolve the MRI. Converting between
coordinate and inertial quantities, $\tau_{_\mathrm{MRI}}$ is changed by
the same factor as $\Omega^{-1}$. The first estimate of
Eq.~\eqref{eq:tau_lambda} is, therefore, preserved as is.

\begin{figure}
  \begin{center}
     \includegraphics[angle=0,width=9.0cm]{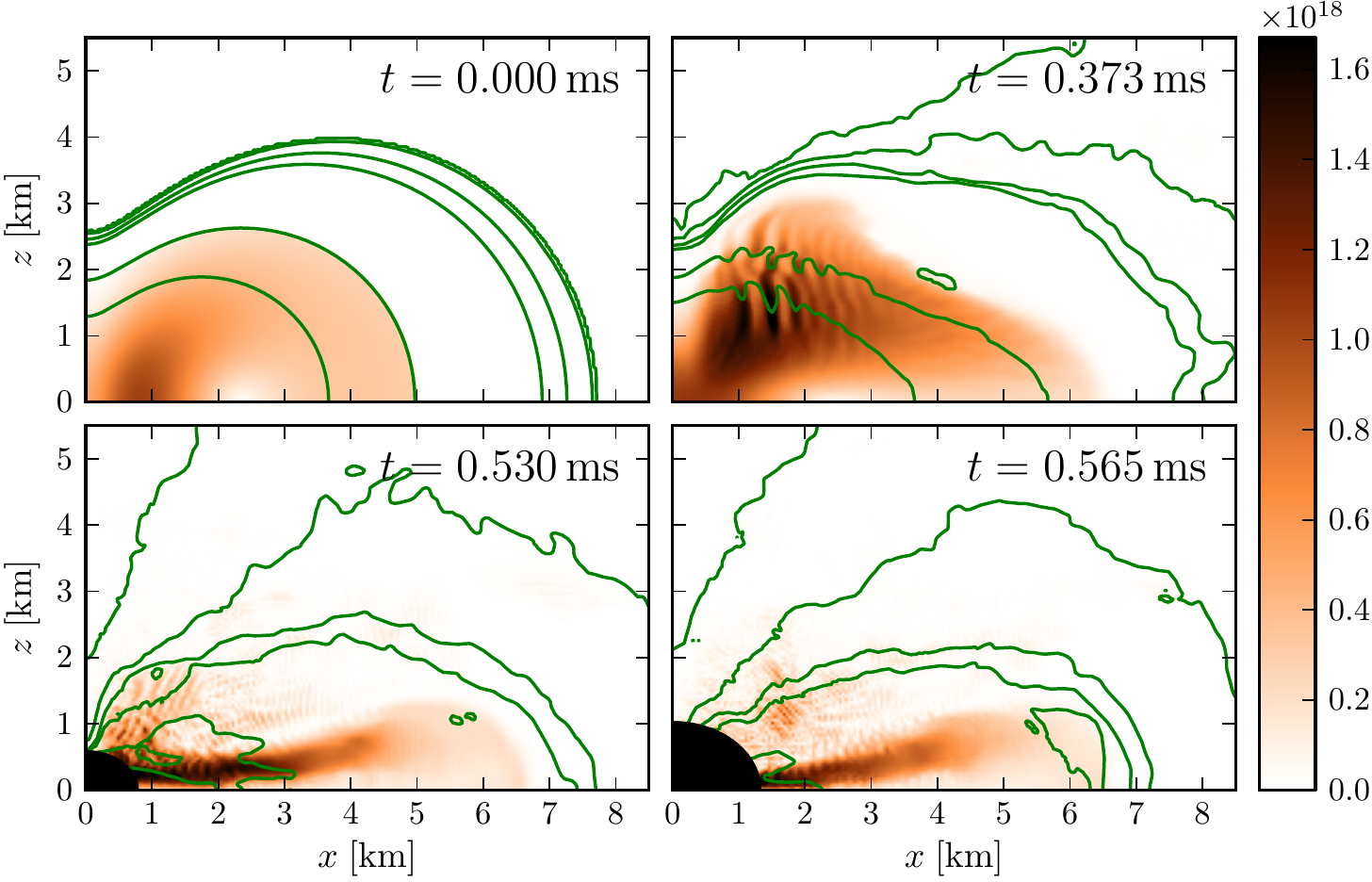}
  \end{center} 
  \caption{Rest-mass density contours ($\rho=10^j\,\mathrm{g/cm}^3$ with
    $j=15.7,\,15,\,14.7,\,14,\, 13.7,\,12.7$, and $11$) and the norm of
    the magnetic field in G in the $(x,z)$ plane at four representative
    times. The region inside the horizon is masked for reasons of clarity.}
\label{fig:4panels}
\end{figure}

\vspace{-0.2cm}
\section{Numerical results}
\vspace{-0.3cm}
Figure~\ref{fig:4panels} shows a
section in the $(x,z)$ plane for our fiducial simulation (i.e., with
$B_c^\mathrm{in}=5\times 10^{17}$~G) in terms of the color-coded norm of
the magnetic field and selected density contours for four characteristic
stages of the evolution. These are: the initial configuration, which
shows a highly flattened HMNS due to rapid rotation; the stage of
pronounced MRI development indicated by the ripples in the magnetic field
and the rest-mass density; the time of collapse to a black hole, when the
apparent horizon is formed; the early post-collapse phase with a
magnetized and geometrically thick torus being formed in the vicinity of
the black hole. We concentrate here only on the MRI in the interior of
HMNSs, leaving the discussion of the potential development of the MRI in
the torus to Refs.~\cite{Rezzolla:2011,Etienne2012b} and to future work.

\begin{figure}
  \begin{center}
     \includegraphics[angle=0,width=8.0cm]{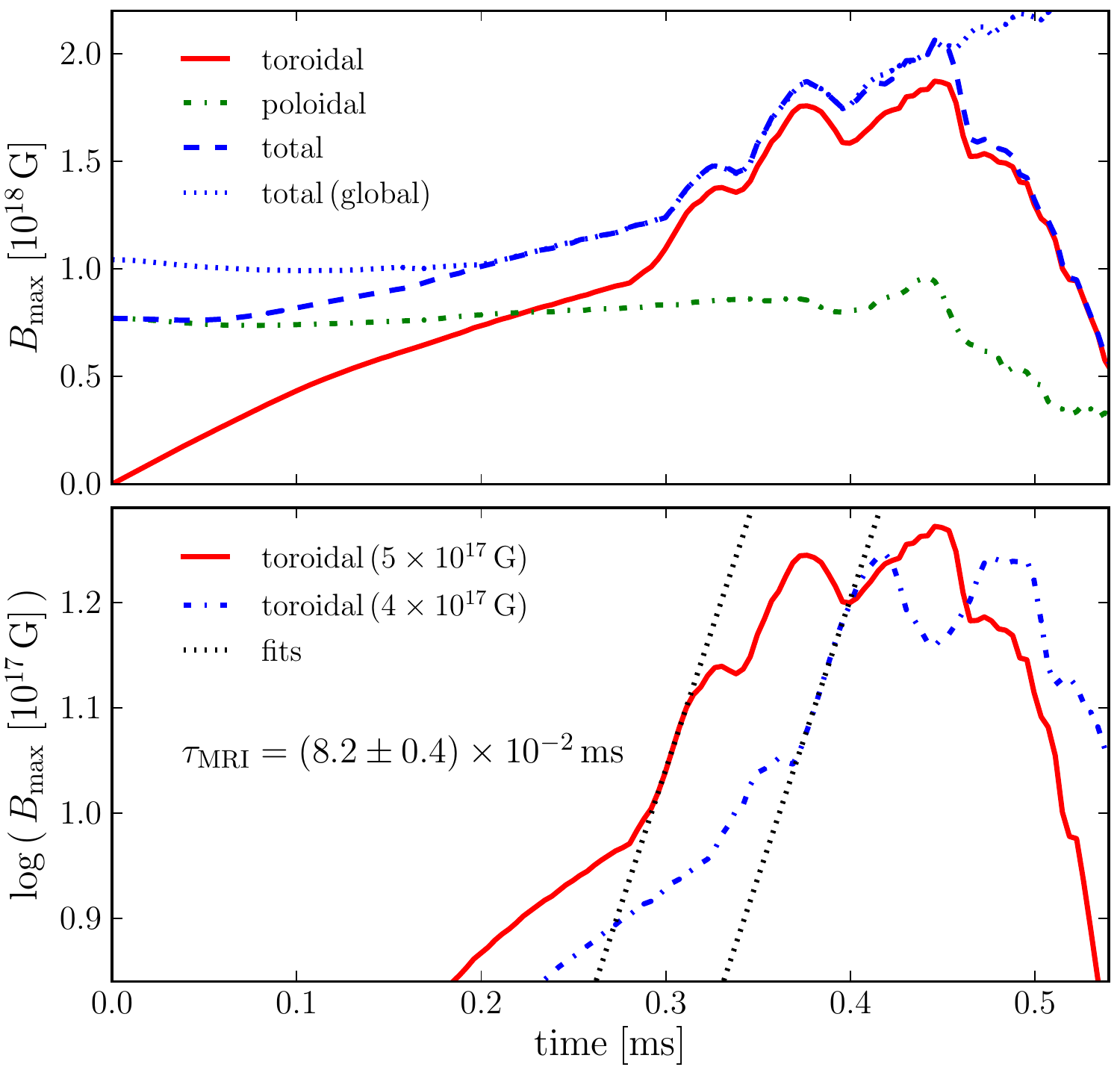}
  \end{center} 
   \caption{Top panel: Evolution of maximum toroidal, poloidal
     and total magnetic fields in a selected region (see text) and of
     the maximum total field in the full computational domain
     (global), for $B_c^\mathrm{in}=5\times 10^{17}$\,G. Bottom
       panel: Toroidal field evolution in log scale for
     $B_c^\mathrm{in}=4$ and $5\times 10^{17}$\,G. The dotted lines
     represent fits to the exponential growths with identical associated
     growth times $\tau_{_\mathrm{MRI}}$.}
\label{fig:MRI}
\end{figure}

\begin{figure}[t]
  \begin{center}
     \includegraphics[angle=0,width=8.0cm]{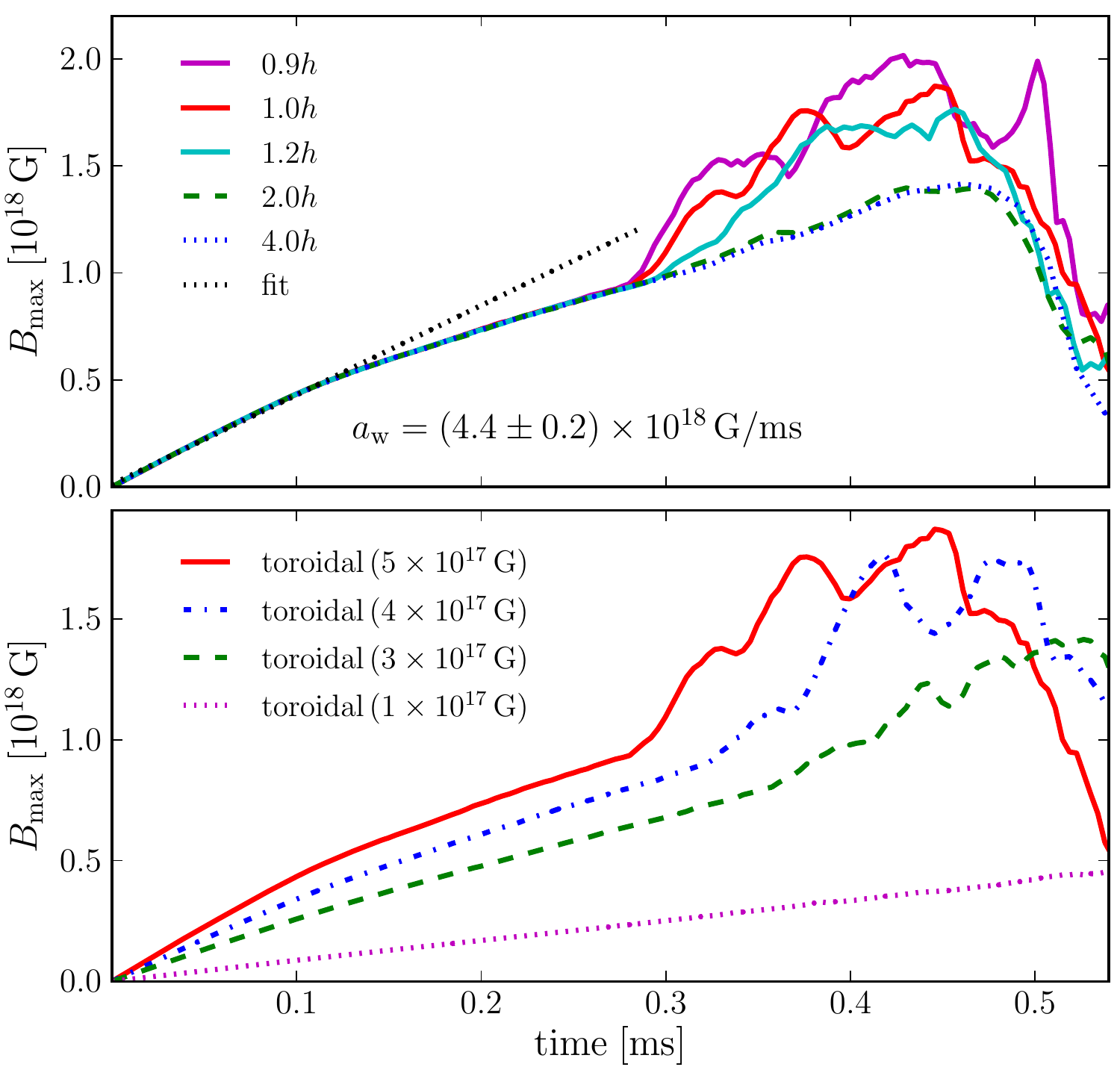}
  \end{center} 
  \caption{Top panel: Evolution of the maximum toroidal magnetic
    field in a selected region for $B_c^\mathrm{in}=5\times 10^{17}$\,G
    and different resolutions $(0.9-4.0)h$, where $h$ is 44\,m. The dash-dotted straight
    line is a fit to the initial linear growth stage of magnetic
    winding, common to all the simulations. Bottom panel: Same as top panel for fixed resolution
    $h$, but for different values of $B_c^\mathrm{in}$.}
\label{fig:NOMRI}
\end{figure}

\begin{figure}
  \begin{center}
     \includegraphics[angle=0,width=8.0cm]{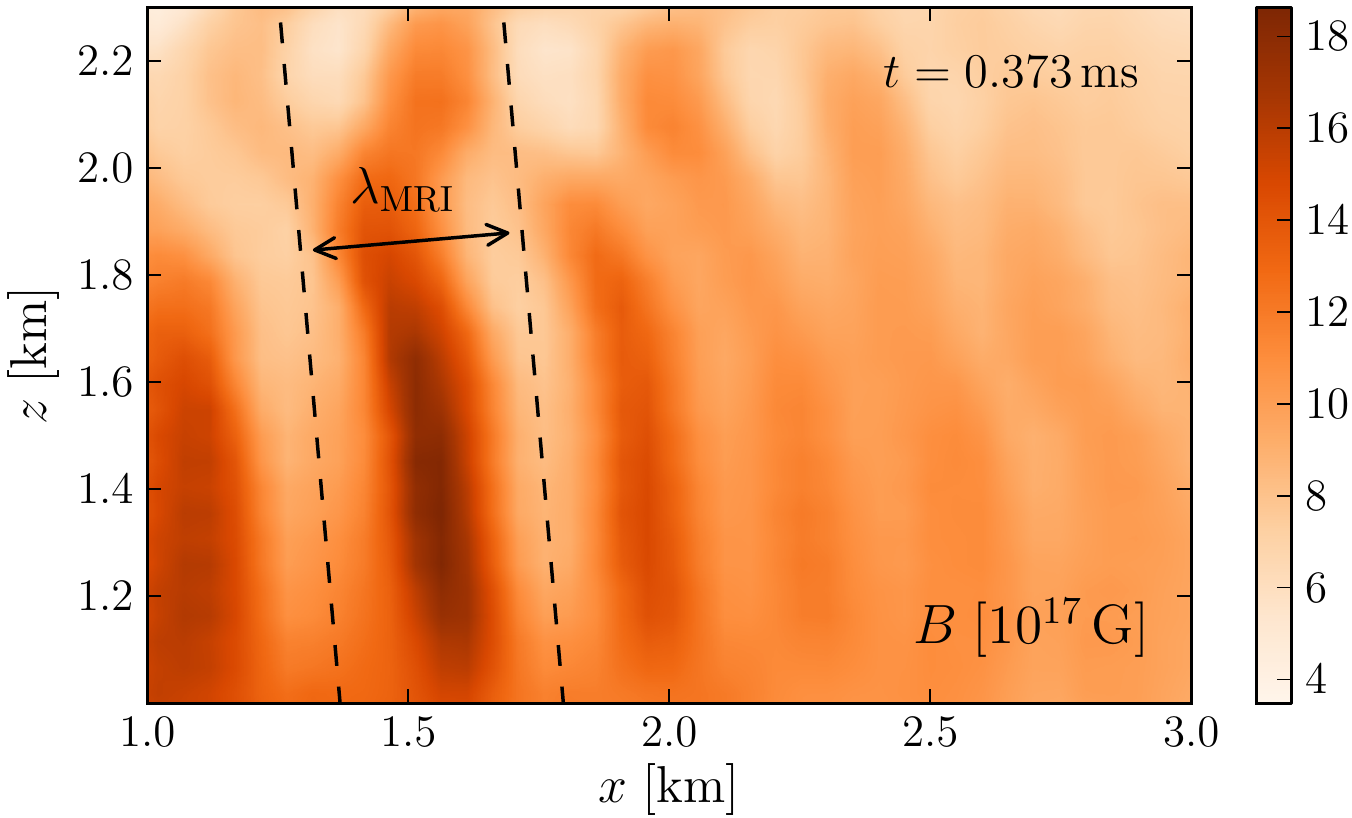}
     \includegraphics[angle=0,width=8.0cm,height=4.125cm]{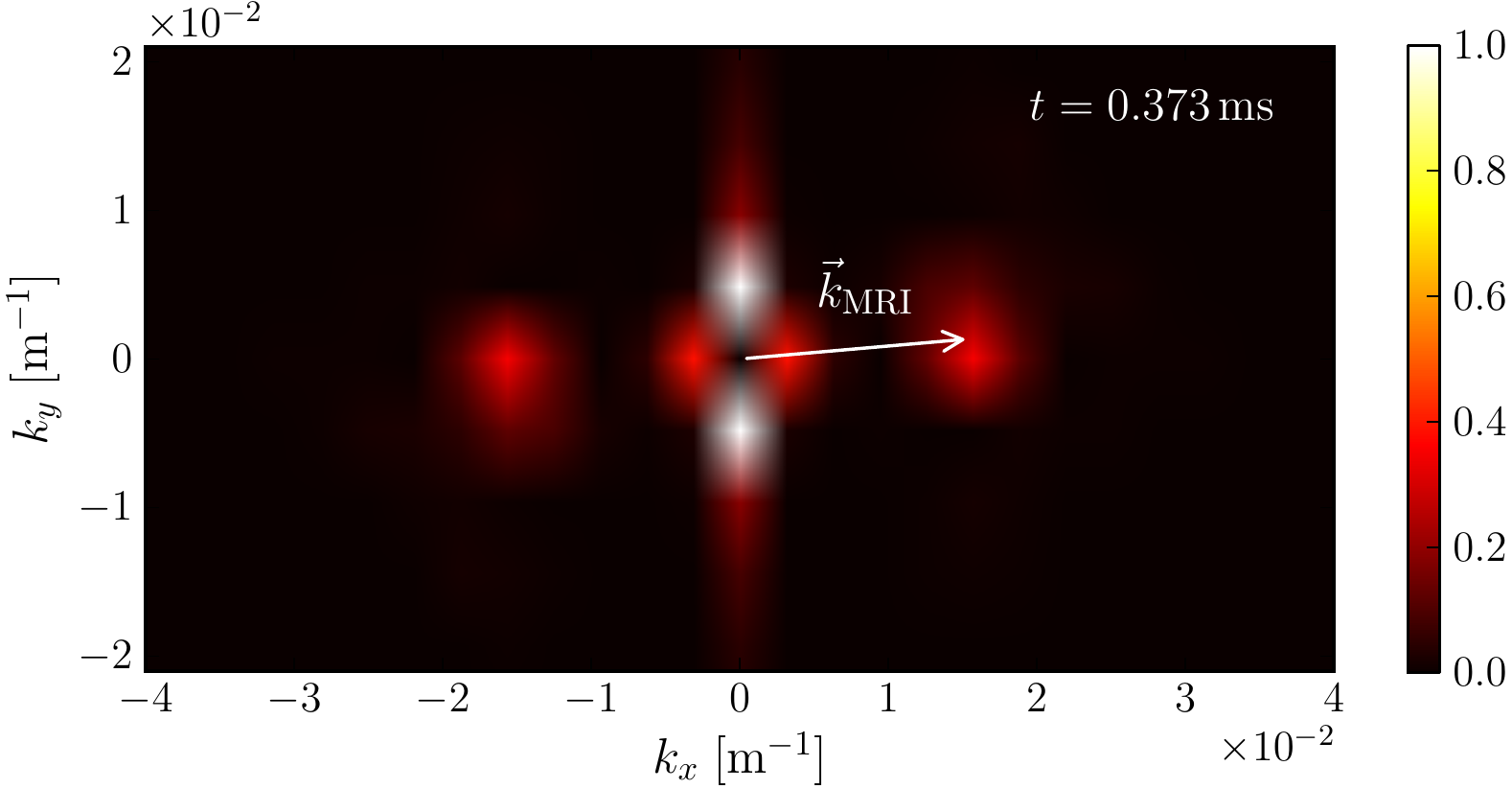}
  \end{center} 
  \caption{Top panel: Norm of the total magnetic field in the
    selected region showing the fastest-growing MRI mode and the onset
    of channel-flow merging (upper part). Bottom panel:
    Corresponding power spectrum showing a single dominant mode with $\lambda_{_\mathrm{MRI}}
    \sim 0.4\,\mathrm{km}$.}
\label{fig:box}
\end{figure}

In order to investigate the properties of the MRI in detail and since the
system at the stage of MRI development is still essentially axisymmetric,
we restrict to a two-dimensional region in the meridional plane defined
by $(x,z)\in[1.0,~3.0] \times[1.0,\,2.3]$\,km, where the MRI is seen most
prominently and which has the typical dimensions of local Newtonian MRI
simulations. In the upper panel of Fig.~\ref{fig:MRI}, we report for our
fiducial model the evolution of the maximum toroidal, poloidal and total
magnetic fields in the selected region until the bulk of the star starts
to collapse and an apparent horizon is formed. The maximum total field in
the full computational domain is also shown, which coincides with the
local one after the magnetic field evolution has become nonlinear around
0.1\,ms. This highlights the fact that the strongest magnetic fields in
the entire computational domain are now to be found inside the selected
region. While the poloidal component of the magnetic field remains
essentially constant up to the collapse, the toroidal component is
significantly amplified during the evolution. This is in contrast with
previous axisymmetric simulations
\cite{Duez:2006qe,Duez2006a}. Initially, the toroidal field shows a
linear growth due to magnetic winding, with a slope
$a_\mathrm{w,\mathrm{fit}}=(4.4\pm 0.2)\times 10^{18}$\,G/ms that matches
the value $a_\mathrm{w}=(4.3\pm 0.7)\times 10^{18}$\,G/ms obtained by
averaging the prediction of Eq.~\eqref{eq:winding} in the region of
interest (see also Fig.~\ref{fig:NOMRI}, upper panel). After $\sim
0.3$\,ms, we distinguish two stages of exponential magnetic-field growth
which coincide with the appearance of coherent channel-flow structures in
the total magnetic field strength (the ``ripples'' in the top right panel
of Fig.~\ref{fig:4panels} and the top panel of Fig.~\ref{fig:box}). These
are the characteristic signatures of the MRI found in local Newtonian
axisymmetric simulations \cite{Obergaulinger:2009}. Note that the
intermediate phase between the two growth periods coincides with the
rearrangement of channel-flow structures. This can be seen in the upper
portion of the upper panel of Fig.~\ref{fig:box}, and is reminiscent of
the channel flow merging reported in \cite{Obergaulinger:2009} (see also
\cite{Cerda2008}). Growth times $\tau_{_\mathrm{MRI}}$ associated with
exponential rises in the toroidal field have been extracted for two
different initial magnetic field strengths (cf. Fig.~\ref{fig:MRI}, lower
panel). The values resulting from both fits agree within error bars and
give $\tau_{_{\mathrm{MRI}},{\rm fit}} = (8.2 \pm 0.4)\times
10^{-2}\,\textrm{ms}$ (the error bar refers to the error from the
fit). This is consistent with the analytic expectation that
$\tau_{_\mathrm{MRI}}$ should be independent of the magnetic field
strength. Furthermore, $\tau_{_{\mathrm{MRI}},{\rm fit}}$ is also in
reasonable agreement with the values predicted by
Eq.~\eqref{eq:tau_lambda} for the selected region: $\tau_{_\mathrm{MRI}}
\approx (4-5)\times 10^{-2}\,\textrm{ms}$.

Figure~\ref{fig:NOMRI} verifies additional important features of the
MRI. The upper panel presents the maximum toroidal magnetic field in the
selected region for the same initial data (with $B_c^\mathrm{in}=5\times
10^{17}$\,G) evolved using five grid resolutions ranging from $0.9h -
4.0h$ (with $h$ referring to the fiducial grid spacing of 44\,m). For the
two coarsest resolution runs ($2h,\,4h$), there are fewer than five grid
points per $\lambda_{_\mathrm{MRI}}$ (see below). The MRI, therefore,
cannot be resolved in these cases. Increasing the resolution, we
gradually recover the growth rate of the fiducial simulation. For the two
finest resolutions ($0.9h$, $1.0h$), the extracted growth rates agree
within error bars. Note that small differences in the maximum magnetic
field after the rapid growth periods are expected when the resolution is
changed. This is because with higher resolution we capture also smaller
wavelengths, which couple nonlinearly and lead to slightly different
magnetic-field amplifications. All of our runs recover the same expected
magnetic winding behavior in the initial phase of the evolution.

The lower panel of Fig.~\ref{fig:NOMRI} illustrates the effect of
varying the initial magnetic field strength at fixed grid resolution
$h$. It validates the disappearance of the MRI when
$\lambda_{_\mathrm{MRI}}$ becomes too small compared with the
resolution. Since $\lambda_{_\mathrm{MRI}}\propto k_{_\mathrm{MRI}}^i
B_i\propto B_{\mathrm{pol}}$ and the poloidal field strength
$B_{\mathrm{pol}}$ remains approximately constant even during the MRI
development (cf. upper panel of Fig.~\ref{fig:MRI}), the number of grid
points per $\lambda_{_\mathrm{MRI}}$ decreases as the initial magnetic
field strength is lowered. At some point, the MRI can no longer be
resolved. We detect a well-resolved instability only when
$B_c^\mathrm{in}> 3\times 10^{17}$~G. The lower panel of
Fig.~\ref{fig:NOMRI} also illustrates that increasing the initial
magnetic field strength decreases the HMNS lifetime (this amounts to a
factor $\gtrsim 2$ with respect to the nonmagnetized case). This is due
to more efficient outward transport of angular momentum which reduces the
centrifugal support in the HMNS~\cite{Giacomazzo:2010}.

The upper panel of Fig.~\ref{fig:box} is a typical snapshot of the norm
of the magnetic field in the selected region after the MRI has fully
developed ($t=0.373\,\mathrm{ms}$). It illustrates the characteristic
coherent channel-flow structures of the instability, which have not been
observed in previous HMNS simulations, nor in global three-dimensional
general-relativistic simulations. Note that such structures are observed
not only for the norm of the magnetic field, but also, e.g., in the
toroidal velocity and magnetic field. The clarity with which these
structures emerge allows us to directly measure the wavelength of the
fastest growing mode. The corresponding two-dimensional power spectrum is
depicted in the lower panel of Fig.~\ref{fig:box}, which---apart from
the maxima around the origin representing large-scale gradients over the
selected region---clearly shows the presence of a single dominant mode
$k_{_\mathrm{MRI}}^i$ nearly parallel to the $x$ axis and corresponding
to a wavelength of $\lambda_{_\mathrm{MRI}}\approx
0.4\,\mathrm{km}\approx 9h$. Note that this geometry is different from
the most commonly considered MRI scenarios where $k_{_\mathrm{MRI}}^i$ is
aligned with the spin axis. There is not enough resolution in the Fourier
domain to accurately measure the very small angle $\theta_{kx}$ between
$k_{_\mathrm{MRI}}^i$ and the $x$ axis, which varies slightly with
time ($\theta_{kx}\approx 3^\circ - 7^\circ$). Using this range of values
for $\theta_{kx}$, the wavelength predicted by Eq.~\eqref{eq:tau_lambda}
for the region of interest is $\lambda_{_\mathrm{MRI}} \approx
(0.5-1.5)$~km, which is in good agreement with the measured value. It
should be emphasized that the analytical estimates of
Eq.~\eqref{eq:tau_lambda} arise from a number of simplifying
assumptions, such as Newtonian physics, axisymmetry, near equilibrium, and the
short-wavelength approximation. None of these assumptions are strictly
valid in our simulations. Notwithstanding the good agreement between our
measurements and Eq.~\eqref{eq:tau_lambda}, a better analytic description
of the MRI is needed for relativistic compact objects.

\vspace{-0.23cm}
\section{Conclusions}
\vspace{-0.3cm}
By performing global three-dimensional MHD
simulations of HMNSs, we have observed the emergence of coherent
channel-flow structures which provide direct evidence for the presence
of the MRI in these systems. This is further supported by the
verification of the main properties of the MRI expected from previous
Newtonian analytical and numerical studies in other astrophysical
scenarios. We note that the persistence of these structures is
nontrivial as they may be unstable in three dimensions as a result of
nonaxisymmetric parasitic instabilities of the Kelvin-Helmholtz type
\cite{Goodman1994,Obergaulinger:2009}. 

Showing the presence of the MRI in HMNSs is of great importance as the
instability amplifies magnetic fields exponentially and can thus rapidly
build up the very high magnetic-field strengths needed to launch a
relativistic jet. Our results show that this amplification can already
occur in the precollapse phase without having to wait for an accretion torus to
be formed after black hole creation. The dynamics in the torus can also
amplify magnetic fields efficiently, but at much later times
\cite{Rezzolla:2011}. The amplification of magnetic fields in the HMNS
due to the MRI is less than one order of magnitude in our model. However,
the HMNS considered here is very short lived even in the absence of
magnetic fields~\cite{Giacomazzo:2010}. In longer-lived HMNSs, the MRI
could well reach several $e$-foldings and thus be a key ingredient in
building the physical conditions necessary for launching the relativistic
jet as revealed by the observations of SGRBs.

As a final note of caution, we remark that the system considered here
involves significant idealizations. Assessing how different the
dynamics can be in a HMNS produced via a realistic merger of neutron
stars will be possible only when much larger computational resources
become available. Our expectation is, however, that the qualitative
features of the scenario described here remain under more realistic conditions.

\vspace{-0.2cm}
\section*{Acknowledgments}
\vspace{-0.3cm}
We thank B. Giacomazzo and J.~L. Jaramillo for discussions and W. Kastaun
for help with the visualizations. Support comes through the DFG SFB/Trans-regio 7, ``CompStar'', a Research Networking Programme of the
ESF. R.~C. is supported by the Humboldt Foundation. The calculations have
been performed on the clusters at the AEI.

\bibliographystyle{apsrev4-1-noeprint}
\bibliography{aeireferences}

\end{document}